\documentclass[11pt]{article}

\usepackage{epsfig}
\usepackage{amssymb}
\usepackage{graphicx}
\usepackage{color}
\usepackage{subfigure}

\begin{document}

\baselineskip 6mm
\renewcommand{\thefootnote}{\fnsymbol{footnote}}


\newcommand{\nc}{\newcommand}
\newcommand{\rnc}{\renewcommand}



\newcommand{\tcb}{\textcolor{blue}}
\newcommand{\tcr}{\textcolor{red}}
\newcommand{\tcg}{\textcolor{green}}


\def\be{\begin{equation}}
\def\ee{\end{equation}}
\def\ba{\begin{array}}
\def\ea{\end{array}}
\def\bea{\begin{eqnarray}}
\def\eea{\end{eqnarray}}
\def\nn{\nonumber\\}


\def\ct{\cite}
\def\la{\label}
\def\eq#1{(\ref{#1})}


\def\a{\alpha}
\def\b{\beta}
\def\g{\gamma}
\def\G{\Gamma}
\def\d{\delta}
\def\D{\Delta}
\def\ep{\epsilon}
\def\e{\eta}
\def\ph{\phi}
\def\Ph{\Phi}
\def\ps{\psi}
\def\Ps{\Psi}
\def\k{\kappa}
\def\l{\lambda}
\def\L{\Lambda}
\def\m{\mu}
\def\n{\nu}
\def\th{\theta}
\def\Th{\Theta}
\def\r{\rho}
\def\s{\sigma}
\def\S{\Sigma}
\def\ta{\tau}
\def\o{\omega}
\def\O{\Omega}
\def\pr{\prime}


\def\half{\frac{1}{2}}

\def\goto{\rightarrow}

\def\na{\nabla}
\def\grad{\nabla}
\def\curl{\nabla\times}
\def\div{\nabla\cdot}
\def\pa{\partial}

\def\bra{\left\langle}
\def\ket{\right\rangle}
\def\lb{\left[}
\def\lc{\left\{}
\def\ls{\left(}
\def\lp{\left.}
\def\rp{\right.}
\def\rb{\right]}
\def\rc{\right\}}
\def\rs{\right)}

\def\vac#1{\mid #1 \rangle}


\def\td#1{\tilde{#1}}
\def\check{ \maltese {\bf Check!}}


\def\Tr{{\rm Tr}\,}
\def\det{{\rm det}}


\def\bc#1{\nnindent {\bf $\bullet$ #1} \\ }
\def\ch {$<Check!>$ }
\def\ss {\vspace{1.5cm}}

\begin{titlepage}

\hfill\parbox{5cm} { }

\vspace{25mm}

\begin{center}
{\Large \bf  Non-conformal Hydrodynamics in Einstein-dilaton Theory}

\vskip 1. cm
  {  Shailesh Kulkarni$^a$\footnote{e-mail : skulkarnig@gmail.com},
  Bum-Hoon Lee$^{ab}$\footnote{e-mail : bhl@sogang.ac.kr},
  Chanyong Park$^a$\footnote{e-mail : cyong21@sogang.ac.kr},
  and Raju Roychowdhury$^a$\footnote{e-mail : raju.roychowdhury@gmail.com}
  }

\vskip 0.5cm

{\it $^a\,$Center for Quantum Spacetime (CQUeST), Sogang University, Seoul 121-742, Korea}\\
{\it $^b\,$Department of Physics,, Sogang University, Seoul 121-742, Korea}\\

\end{center}

\thispagestyle{empty}

\vskip2cm


\centerline{\bf ABSTRACT} \vskip 4mm

\vspace{1cm}

In the Einestein-dilaton theory with a Liouville potential parameterized by $\eta$, we find
a Schwarzschild-type black hole solution. This black hole solution, whose asymptotic geometry 
is described by the warped metric, is thermodynamically stable only for $0 \le \eta < 2$.
Applying the gauge/gravity duality, we find that the dual gauge theory represents a non-conformal 
thermal system with the equation of state depending on $\eta$.
After turning on the bulk vector fluctuations with and without a dilaton coupling, 
we calculate the charge diffusion constant, which indicates that 
the life time of the quasi normal mode decreases with $\eta$. 
Interestingly, the vector fluctuation with the dilaton coupling shows that the DC conductivity
increases with temperature, a feature commonly found in electrolytes.
\vspace{2cm}


\end{titlepage}

\renewcommand{\thefootnote}{\arabic{footnote}}
\setcounter{footnote}{0}

\tableofcontents

\section{Introduction}
In recent years, the applications of AdS/CFT correspondence \cite{maldacena,klebanov,witten} to 
strongly interacting gauge theories is one of the active frontiers in string theory.  
The more general concept, the so called gauge/gravity duality, has been widely adopted to get a 
better understanding of QCD and condensed matter systems in the strong coupling regime, on a non-AdS 
space like Lifshitz, Schr\"{o}dinger  and anisotropic space as dual geometries \cite{Kachru:2008yh}-\cite{Lee:2011zzf}. 
This correspondence ushered in a fresh hope to test general lore about the quantum field theory in a
non-perturbative setting and thus learn general lessons regarding the strongly coupled dynamics.

In \cite{Iqbal:2008by,Kanitscheider:2008kd,Perlmutter:2010qu}, a very general black brane geometry, which also includes an asymptotically 
non-AdS space, was examined to study the linear response transport coefficients of a strongly 
interacting theory at finite temperature in the hydrodynamic limit. Authors showed that,
there exists the universality of the shear viscosity in terms of the universality of the 
gravitational coupling. In addition, they have also clarified the relation between the transport 
coefficients at the horizon and boundary by using the holographic renormalization flow 
\cite{Iqbal:2008by,Skenderis:2002wp}.

In the AdS black brane geometry the asymptotic limit is usually identified with an
UV fixed point of the gauge theory.  The warped geometry, one of the non-AdS examples, lacks this feature. 
What is the  meaning of the asymptotic non-AdS space from the point of view of the gauge/gravity duality?
If we add a relevant or marginal operator to the conformal field theory at the UV fixed point,
the conformality is still preserved at least at that point \cite{Goldstein:2009cv,Goldstein:2010aw}. However,
an irrelevant operator can break the conformality of the theory away from the fixed point.  
This gives rise to the non-AdS geometry in the bulk. Thus, the warped geometry might describe such a deformed  
conformal field theory with an inclusion of an irrelevant operator.

Another interesting application of gauge/gravity duality can be found in the arena of AdS/CMT 
\cite{Sachdev}-\cite{Liu} and the fluid/gravity correspondence \cite{Janik:2005zt}-\cite{Hubeny:2011hd}. 
Recently, AdS/CFT correspondence has been widely used to investigate the critical behaviour of the condensed matter systems. 
However, if we are interested in energy scales away from the critical point, we should modify the dual geometry by 
incorporating the dilaton field to describe the running of the gauge coupling in the dual theory. A natural question 
that might crop up is whether there exists an UV fixed point for the condensed matter system.  
Indeed, the existence of a scale associated with the lattice spacing causes the underlying theory to be non-conformal.
This give us a hunch that the warped geometry might be a good candidate to investigate the general lore of 
non-conformal relativistic field theory. Although the warped geometry is not yet treated on the same footing as 
that of the AdS spacetimes, it is worthwhile to explore up to what extent the usual AdS/CFT techniques
can be stretched. 

Motivated by this, we shall study in some detail the Einstein-dilaton theory with a 
Liouville potential. Generally a Liouville potential is used to give mass to the dilaton, 
as can be derived from higher dimensional string theory with a deficit central charge 
(see for example \cite{Harvey93, Horowitz93}). The physical implications for
considering a Liouville potential in the action was studied in great depth in the 
context of Einstein-Maxwell-dilaton theory in \cite{Charmousis09}. With a Liouville term the asymptotic 
structure gets modified and no more do we get an asymptotically AdS solution, rather we have an warped geometry. 
The isometry group of a warped geometry is usually smaller than that of the AdS space, however, it still preserves the 
Poincare symmetry at the asymptotic region. 
This fact implies that if there exists a dual gauge theory in the sense of the gauge/gravity duality, 
the corresponding gauge theory should be relativistic and non-conformal. The non-conformality is closely 
related to the non-trivial profile of the bulk dilaton field.
 
In this paper,  we consider Einstein-dilaton theory with a Liouville potential parametrized by $\eta$. 
By solving the Einstein as well as the dilaton equations simultaneously, we are able to find a 
black hole solution with an warped asymptotic geometry. This black hole satisfies the laws of 
thermodynamics \cite{Bek1,Hawk3} and represents a thermally equilibrated system whose equation of state 
parameter $w$ crucially depends on the parameter of the theory $\eta$. For $\eta = 0$, 
our black hole geometry reduces to the usual AdS Schwarzschild black hole and represents an equilibrium 
system with conformal matter (energy-momentum tensor being traceless). Next, employing the gauge/gravity duality 
to our warped geometry, we obtain a dual gauge theory with non-conformal matter.
We will follow a similar technique as prescribed in \cite{Son:2002sd,Policastro:2002se,Policastro:2002tn}.
 In order to get a better handle over the non-conformal theory, we further investigate the charge dissipation 
 in the hydrodynamic limit. We achieve this by turning on the bulk vector fluctuations with and without a dilaton 
 coupling. Through these investigations, we find that the non-conformality increases the charge diffusion constant 
 and thus shortens the life time of the corresponding quasi normal mode. In absence of the dilaton coupling, 
 the real conductivity does not depend on the non-conformality. If we consider the vector fluctuations coupled to 
 the dilaton field, the conductivity of the dual non-conformal theory  depends on temperature. More precisely, 
 the conductivity increases with temperature. This type of behaviour is commonly found in electrolytes.
Therefore, it is worthwhile to investigate the physical properties of such thermodynamic systems at length and 
compare it with available data from the condensed matter physics. 

The plan of this paper is as follows: In Sec.2, we discuss the black brane solution of 
the Einstein-dilaton theory and the corresponding thermodynamics. 
Taking into account the gauge/gravity duality in this setting, we find that the dual field theory 
should be described in terms of a non-conformal gauge theory.  
Sec.3 is devoted to the computation of gauge fluctuations without the dilaton coupling. From this, 
we investigate the hydrodynamic transport coefficients viz. the charge diffusion constant and 
conductivity of the non-conformal dual gauge theory. In Sec. 4, we redo similar sort of calculation 
for the gauge field fluctuation with a dilaton coupling. Here we will find conductivity having a 
nontrivial dependence on the temperature. Finally, we conclude our work with some remarks in 
Sec.5. The Appendices A and B contain the details of the the computation for the longitudinal and 
transverse modes of the vector fluctuation without and with a dilaton coupling.

\section{Einstein-dilaton theory}

We first consider the Einstein dilaton gravity theory
\be
S= \frac{1}{16 \pi G} \int
d^{4}x\sqrt{-g} \lb R- 2 (\pa \phi)^2-V(\ph) \rb,
\ee
with a Liouville-type scalar potential
\be
V(\phi) = 2 \Lambda e^{\eta\phi} ,
\ee
where $\L$ and $\eta$ are the cosmological constant and an arbitrary constant, respectively. Since we are
interested in the AdS-like space, we concentrate on the case having negative cosmological constant,
$\L < 0$ and set $G=1$ for simplicity. Before finding a geometric solution of the above Einstein-dilaton theory, 
we first consider the simplest case. If the scalar field is set to zero, the above action describes the geometry 
with a negative cosmological constant, which is the Anti de-Sitter (AdS) space or Schwarzschild AdS (SAdS) black hole 
(or brane). So, the pure AdS and SAdS black hole are the special solutions of the more general ones.

For $\ph \ne 0$ or $\eta=0$, the Einstein equation and equation of motion for the scalar field are respectively given by
\begin{eqnarray}
R_{\mu\nu}-\frac{1}{2}Rg_{\mu\nu}+\frac{1}{2}g_{\mu\nu}V(\phi) &=& 2 \partial_{\mu}\phi
\partial_{\nu}\phi- g_{\mu\nu}(\pa \phi)^{2} ,  \la{eq:Einstein}\\
\frac{1}{\sqrt{-g}} \partial_{\mu}(\sqrt{-g} g^{\m \n} \partial_{\n}\phi) &=&  \frac{1}{4} \frac{\partial
V(\phi)}{\partial\phi}  \la{eq:scalar}.
\end{eqnarray}
To solve these equations, we take the following metric ansatz
\begin{equation}		\la{eq:asymp}
ds^{2}=-a(r)^{2}dt^{2}+\frac{dr^{2}}{a(r)^{2}}+b(r)^{2}(dx^{2}+dy^{2}) ,
\end{equation}
with
\bea
\phi(r) &=& -k_{0}\log r \nn
a(r) &=& a_{0}r^{a_{1}},~~~b(r)=b_{0}r^{b_{1}}  \la{eq:aandb},
\eea
which were also used for finding the generically hyperscaling violating solutions \cite{Goldstein:2009cv,Charmousis:2010zz,Lee:2010qs,Goldstein:2010aw,Huijse:2011ef,Gouteraux:2011ce,Kanitscheider:2009as,Gouteraux:2011qh}.
By rescaling the $x$ and $y$ coordinates, we can choose $b_0 = 1$ without any loss of 
generality.  In general, since the scalar field should satisfy the second order differential equation, the solution of 
the scalar field should involve two integration constants. The most general ansatz for the scalar field is
\be \la{eq:dilatonprofile}
\ph = \ph_0 - k_0 \log r ,
\ee
where $\ph_0$ and $k_0$ are two integration constants. However, $\ph_0$ can be set to zero by a suitable rescaling of
the cosmological constant, so we can choose $\ph_0=0$ without any loss of generality.

The non-black hole solution satisfying \eq{eq:Einstein} and \eq{eq:scalar} is given by
\be		\la{sol:nonbh}
a_0 = \frac{(4 + \eta^2) \sqrt{- \L}  }{ 2 \sqrt{12 - \eta^2}} \quad , \quad
k_0 = \frac{2 \eta}{4 + \eta^2} \quad , \quad a_1 = b_1 = \frac{4}{4+\eta^2} ,
\ee
where all constants in the ansatz are exactly determined in terms of  the original parameters, $\eta$ and
$\L$ in the action.

According to the first relation in \eq{sol:nonbh}, the solution is well defined only for
$\eta^2 < 12$, which corresponds to the Gubser bound \cite{Gouteraux:2011ce,Gubser:2000nd,Gursoy:2007cb}. After introducing new coordinates (except for $\eta=0$)
\bea
\td{r} &=& \frac{1}{a_0 ( 1 - a_1 ) } r^{1-a_1} , \nn
\td{t} &=& a_0^{1/(1-a_1)} \ (1-a_1)^{a_1/(1-a_1)} \ t \ , \nn
\td{x} &=&  a_0^{a_1/(1-a_1)} \ (1-a_1)^{a_1/(1-a_1)} \ x \ , \nn
\td{y} &=&  a_0^{a_1/(1-a_1)} \ (1-a_1)^{a_1/(1-a_1)} \ y \ ,
\eea
the metric solution can be rewritten in the form of an warped geometry
\be		\la{met:asymp}
ds^2 = \td{r}^{2 a_1 /(1 - a_1)} \lb - d \td{t}^2 + d \td{x}^2 + d \td{y}^2 \rb + d \td{r}^2 .
\ee
Notice that the asymptotic warped geometry does not reduce to the AdS space for 
$\eta \ne 0$ but preserves the ISO(1,2) isometry. Especially, for $\eta=0$ together with $\L=-3$, the above metric in 
\eq{eq:asymp} reduces to the usual AdS metric without a dilaton field, in which the isometry group is enhanced to SO(2,3) 
corresponding to the conformal group of the dual gauge theory. For $\eta \ne 0$, if we assume that the gauge/gravity duality 
is still working, the dual gauge theory is not conformal but still contains the $2+1$-dimensional Poincare symmetry ISO(1,2), 
which corresponds to the relativistic non-conformal matter theory.
Although the dual theory of this background is non-conformal, if we consider the uplifting
of it to a higher dimension, the conformal symmetry can be restored
\ct{Huijse:2011ef,Gouteraux:2011ce,Kanitscheider:2009as,Gouteraux:2011qh}.  
\footnote{We would like to thank E. Kiritsis for pointing this to us.}

The warped geometry can be easily generalized to the black hole geometry. Since the black hole solution is exactly the same as 
the black brane solution in the Poincare patch, we will concentrate on the Poincare patch without distinguishing them from now on.
For the black hole, we consider a slightly different metric ansatz 
\cite{Chamblin:1999ya,Goldstein:2009cv,Charmousis:2010zz,Cadoni:2011nq}  
\begin{equation}		\la{ansatz}
ds^{2}=-a(r)^{2} f(r) dt^{2}+\frac{dr^{2}}{a(r)^{2} f(r)}+b(r)^{2}(dx^{2}+dy^{2}) ,
\end{equation}
with the following black hole factor 
\be
f(r) = 1 - \d \ m \ r^{-c} ,
\ee
where $m$ is the black hole mass and a constant $\d$ is introduced for later convenience
\be
\d = \frac{8 \pi}{V_2 (-\L)} \frac{12-\eta^2}{4+\eta^2}.
\ee
Here,  $V_2 = \int_0^{L} dx dy$ is a regularized area in $(x,y)$ plane with an appropriate infrared cutoff $L$.
Then, the solution of the Einstein-dilaton theory is described by the same constants in \eq{sol:nonbh} along with
\be
c = \frac{12-\eta^2}{4 + \eta^2} .
\ee
Notice that since $\eta^2 < 12$, $c$ is always positive. In the asymptotic region $r \to \infty$,
the metric reduces to the previous one. This metric contains the effect of the scalar field. Strictly speaking, 
 this corresponds to a black brane due to the translational symmetry in $x$- and $y$-directions.

From the metric of the uncharged black hole, we can easily find the horizon $r_h$ satisfying $f (r_h) = 0$
\be
m = \frac{r_h^{(12-\eta^2)/(4+ \eta^2)}}{\d} .
\ee
The Hawking temperature $T_H$ defined by the surface gravity at the horizon, is given by
\bea 		  \la{eq:BHtemperature}
T_H &\equiv& \frac{1}{4 \pi} \frac{\pa}{\pa r} \lc a(r)^2 f(r) \rc |_{r=r_h} \nn
      &=& \frac{(- \L) (4 + \eta^2)}{16 \pi} \ r_h^{(4-\eta^2)/(4+\eta^2)} .
\eea
The Bekenstein-Hawking entropy $S_{BH}$ is
\bea
S_{BH} &\equiv& \frac{A(r_h)}{4} \nn
&=& \frac{V_2}{4}  \ r_h^{8/(4+\eta^2)} ,
\eea
where $A(r_h)$ implies the area at the black hole horizon.

Usually, the black hole system provides a well-defined analogous thermodynamic system, so the black hole 
should satisfy the first thermodynamic relation
\be
0 = d E - T_H d S_{BH} .
\ee
From this relation we can determine the energy of the black hole by rewriting the Hawking temperature in terms of the 
Bekenstein-Hawking entropy and then integrating it. In terms of the black hole horizon, the energy is given by
\be
E = \frac{(- \L) V_2}{8 \pi} \frac{4 + \eta^2}{12-\eta^2} \ r_h^{(12-\eta^2)/(4+\eta^2)} ,
\ee
and the free energy of this black hole is given by
\be
F \equiv E - TS = - \frac{(- \L) V_2}{64 \pi} \frac{16 - \eta^4}{12-\eta^2} \ r_h^{(12-\eta^2)/(4+\eta^2)} .
\ee
Following the thermodynamic relation, the pressure of the system is given by $P = -\pa F/\pa V_2$,
so we can easily read off the equation of state parameter from the following relation $P = w E/V_2$
\be
w = \frac{1}{8} (4 -\eta^2) .
\ee
Following the gauge/gravity duality, we can interpret the thermodynamic quantities of the black hole system as ones of the dual 
gauge theory defined on the boundary. For $\eta=0$, the solution of gravity theory is given by the AdS black hole and corresponds 
to the gauge theory with the conformal matter whose energy-momentum tensor is traceless. Since $0< \eta^2 < 12$, 
the equation of state parameter $\o$ can have the following values
\be 
-1 < w < \half ,
\ee
which corresponds to the gauge theory with the non-conformal matter.

To check the thermodynamic stability of the dual gauge theory including non-conformal matter,  we calculate the specific 
heat of the black hole
\bea  \la{sp heat:uncharged hairy}
C_{uh} &\equiv& \frac{d E}{d T_H} \nn
&=& \frac{(- \L) V_2}{8 \pi } \frac{4 + \eta^2}{4-\eta^2} \ls \frac{16 \pi}{(- \L) (4 + \eta^2)} \rs^{(12-\eta^2)/(4-\eta^2)} 
T_H^{8/(4-\eta^2)} .
\eea
The specific heat of the SAdS black hole can be obtained by setting $\eta=0$ and it is always positive, which implies that the 
SAdS black hole is thermodynamically stable.
There exists another critical point called the crossover point $\eta^2=4$ \cite{Gouteraux:2011ce,Gursoy:2007cb}. For the Einstein-Maxwell-dilaton theory, 
the critical points, the Gubser bound and the crossover value for finite density geometries were studied in depth in \cite{Charmousis:2010zz}.
 For $\eta^2 < 4$, the black hole has positive specific heat.
For $\eta^2=4$, the specific heat of the black hole is singular, while, in the other parameter region $4 < \eta^2 < 12$, it is negative.
As a result, the  black hole is stable only for $0 \le \eta^2 < 4$, which implies, from the 
dual gauge theory point of view, that the non-conformal matter having the equation of state parameter in the
following region $0 < w <  \half$ provides the thermodynamically stable system. In other cases $-1 < w < 0$, the dual gauge theory is thermodynamically unstable.

\section{Properties of the non-conformal dual gauge theory}

In the previous section, we have shown that if we use the gauge/gravity duality in the non-AdS 
background, then the warped geometry, which is the solution of the Einstein-dilaton theory, can 
describe the non-conformal dual gauge theory. To understand the physical properties of this 
non-conformal dual gauge theory, we need to investigate the linear response of the vector 
fluctuation in the hydrodynamic limit with small frequency and momentum.
Furthermore, the 4-dimensional warped geometry obtained here may originate from the 10 -
dimensional string theory and depending on the compactification mechanism we can treat various 
vector fluctuations either with or without the dilaton coupling. In the gauge/gravity duality, the 
nontrivial dilaton profile can be identified with the running coupling constant of the dual gauge theory.
It was also shown that the nontrivial dilaton coupling of the gauge fluctuation plays an 
important role to determine the properties of the  dual gauge theory leading to the strange metallic 
behavior 
\ct{Goldstein:2009cv,Charmousis:2010zz,Lee:2010qs,Lee:2011zzf,Karch:2007pd,Hartnoll:2009ns,Koutsoumbas:2009pa}. Therefore it is, 
indeed interesting to study hydrodynamic properties of the vector 
fluctuations with or without a nontrivial dialton profile. In this and the next section, we will investigate such hydrodynamic 
properties without and with a nontrivial dilaton coupling, respectively.

\subsection{Vector fluctuation without the dilaton coupling}

We consider Maxwell field action as a fluctuation over the background geometry \eq{ansatz},
\be
S_{M} = -\frac{1}{4 g_4^2}\int d^{4}{x} \sqrt{-g} F^{\m\n}F_{\m\n}, \la{eq:maxwellaction}
\ee
where
\be
F_{\m\n} = \pa_{\m}A_{\n} - \pa_{\n}A_{\m} ,
\ee
and $g_4^2$ is a constant coupling of the bulk gauge field.
In the $A_{r}=0$ gauge, we take $A_{i}$  ($i = t,x,y$) in the Fourier space as
\be
A_{i}(t,{\bf x},r) = \int \frac{d\o d^{2}{q}}{(2\pi)^3} e^{-i(\o t - \bf {q\cdot x})}
 A_{i}(\o,{\bf q},r) ,
 \la{eq:gaugefourier}
\ee
where ${\bf q}$ = $\ls q_{x},q_{y} \rs$ and ${\bf x}= \ls x,y \rs$. Due to the rotation symmetry along the
$\ls x,y \rs$ plane,
we can consider, without any loss of generality, the momentum only along $y$ direction like
${\bf q}= \ls 0, q \rs$.
Then, the equations of motion for the gauge fluctuations
\be
\pa_{\n}\lb\sqrt{-g} g^{\n\rho}g^{\m\sigma}F_{\sigma\rho}\rb =0 \la{eq:gaugefluct},
\ee
can be reduced to two parts, longitudinal and transverse.
The longitudinal part is given by the following set of coupled equations for $A_t$ and $A_y$
\bea
0 &=&b^2 \o A'_{t} + g q A'_{y} \la{eq:Arr}\\
0 &=&b^2 A''_{t} +2bb' A'_{t}  - \frac{1}{g}(q\o A_{y}+ q^2 A_{t})\la{eq:Atr}\\
0 &=&g A''_{y} + g'A'_{y} + \frac{1}{g}(\o q A_{t}+\o^{2}A_{y})  \la{eq:Ayr} ,
\eea
where the prime denotes the derivative with respect to the radial coordinate $r$ and
new function $g(r)$ is introduced for later convenience
\be
g(r) = a(r)^2 f (r) .
\ee
The transverse mode governed by $A_x$ satisfies the following equation
\be
0 = A''_{x} + \frac{g'}{g}A'_{x} + \frac{1}{g^2}\lb\o^2 -q^2\frac{g}{b^2}\rb A_{x} .
\la{eq:Axr}
\ee

To solve the coupled equations we introduce new coordinate $z$
\be
z = \frac{T_{H}}{\tilde{\L}} \frac{r^{1-2a_{1}}}{2a_{1}-1} ,\la{eq:zcoordinate}
\ee
with
\be
\tilde{\L}= \ls\frac{-\L}{16\pi}\rs\frac{(4+\eta^2)^2}{4-\eta^2} , \la{eq:lambdalambdatilde}
\ee
where it must be noted that $\td{\L}$ is independent of temperature. 
Then, the metric components can be rewritten in term of $z$ as
\bea
g(z) &=& a^{2}_{0}\lb z (2a_{1}-1)\ls\frac{\tilde{\L}}{T_{H}} \rs\rb^{e}
\lb 1-z^d\rb  ,\la{eq:fz}\\
b^{2}(z) &=& \lb z (2a_{1}-1)\ls\frac{\tilde{\L}}{T_{H}} \rs\rb^{e} , \la{eq:bu}
\eea
where two constants $d$ and $e$ are defined as
\be
d = \frac{c}{2a_{1}-1} \quad {\rm and}  \quad  e = -\frac{2a_{1}}{2a_{1}-1}. \la{eq:dande}
\ee
Since $0 \le \eta^2 < 4$, we find that $d\ge 3$ and $e \le -2$.  \\
Notice that in the $z$ coordinate the horizon is located at $z=1$ and the asymptotic boundary is
at $z=0$. Using the rescaled frequency and momenta
\be
\tilde{\o} = \frac{\o}{T_{H}}  \quad {\rm and} \quad  \tilde{q} = \frac{q}{T_{H}} , \la{eq:dimlesswq}
\ee
the coupled equations for the longitudinal modes become
\bea
0 &=&\tilde{\o} A'_{t} + F(z) \tilde{q} A'_{y} \la{eq:Arz}\\
0 &=& A''_{t} - \frac{\tilde{\L}^2}{F(z)}\lb \tilde{q}\tilde{\o} A_{y} + \tilde{q}^2 A_{t}\rb ,\la{eq:Atz}\\
0 &=&A''_{y} + \frac{F'(z)}{F(z)}A'_{y} +\frac{\tilde{\L}^2}{F^{2}(z)}\lb \tilde{\o}\tilde{q}A_{t} +
 \tilde{\o}^2 A_{y}\rb , \la{eq:Ayz}
\eea
while the transverse mode is governed by the following,
\be
0 = A''_{x} + \frac{F'(z)}{F(z)} A'_{x} + \frac{\tilde{\L}^2}{F^{2}(z)}\lb \tilde{\o}^2 -
 F(z)\tilde{q}^2\rb , \la{eq:Axz}
\ee
where the prime means the derivative with respect to $z$ and we define
\be
F(z) = \frac{g(z)}{b^{2}(z)} = a^2_{0}(1-z^d) . \la{eq:F}
\ee
At this point we refer our readers to the appendix A for a detailed computation for the longitudinal and transverse modes of vector fluctuation.

\subsection{Charge diffusion constant and Conductivity}

The exposition in Appendix A put us in a position to calculate the retarded Green's functions on the boundary. 
The general strategy to get the Green's function is discussed in \cite{Son:2002sd,Policastro:2002se}.
According to the gauge/gravity duality, the on-shell gravity action corresponding
to the boundary term can be regarded as a generating functional of the dual gauge theory.
Thus, the generating functional of the dual gauge theory can be described by
the boundary action of the bulk gauge fluctuation
\be
S_{B} =  \lp \frac{1}{2 g_4^2}\int dt dx dy \  \sqrt{-g} g^{zz}(z)g^{ij}(z) \ A_{i}(z) \pa_z A_{j}
\right|_{z=0},  \la{eq:boundaryction}
\ee
where $i,j = t,x,y$. Since, the boundary value of the gauge fluctuation plays the role of the source
$A^{0}_{i}$ for an operator  in the dual gauge theory, the retarded Green function of the dual 
operator can be derived by the following relation
\be
\mathcal{G}_{ij}   = \lim_{z\rightarrow 0}\frac{\delta^{2}S_{B}(z)}{\delta A^{0}_{i}\delta A^{0}_{j}}. \la{eq:Greenfndef} 
\ee
The factor $ \sqrt{-g}g^{zz}(z)g^{ij}(z)$ attain a constant value in $z\rightarrow 0$ limit. 
Hence, only the leading terms in  the combination $A_{i}(z)A'_{j}(z)$ contribute to the finite part of the boundary action.
Near the boundary, with the aid of \eq{eq:solnofAtprime}, \eq{eq:solnofAyprime}, \eq{eq:solnofAxz} and \eq{eq:Cx},
we get 
\bea
A_{t}(z)A'_{t}(z) &\approx &  A^{0}_{t} \frac{\ls\tilde{\o}\tilde{q} A^{0}_{y} + \tilde{q}^2 A^{0}_{t}\rs}{ \ls i\frac{\tilde{\o}}{\tilde{\L}}- \tilde{q}^2\rs} + \cdots , \la{eq:boundaryterms}\nn
A_{y}(z)A'_{y}(z) &\approx &  A^{0}_{y}\frac{\ls\tilde{\o}^{2} A^{0}_{y} + \tilde{q}\tilde{\o} A^{0}_{t}\rs}{a^{2}_{0}\ls i\frac{\tilde{\o}}{\tilde{\L}}- \tilde{q}^2\rs}+ \cdots , \nn
A_{x}(z)A'_{x}(z) &\approx &  \ls\frac{\tilde{\L}}{a_{0}}\rs^{2} (A^{0}_{x})^{2} \ls i\frac{\tilde{\o}}{\tilde{\L}}- \tilde{q}^2\rs + \cdots~ ,
\eea
where the ellipsis implies higher order terms which vanish at the boundary. Note that in our case, unlike the 5-dimensional one \cite{Policastro:2002se}, we do not find any divergent terms in the product of $A_{i}(z)$ and $A'_{j}(z)$. 

Using (\ref{eq:boundaryterms}), the boundary action reduces to 
\bea
S_{B} = - \frac{T_{H}}{2 g_4^2}
\lc \frac{A^{0}_{t}(\tilde{q}^{2}A^{0}_{t}+\tilde{\o}\tilde{q}A^{0}_{y})+ A^{0}_{y}(\tilde{\o}^{2}A^{0}_{y}
+ \tilde{\o}\tilde{q})A^{0}_{t}}{i\tilde{\o}-\tilde{\L}\tilde{q}^{2}} -( A^{0}_{x})^{2}(i\tilde{\o}-\tilde{\L}\tilde{q}^{2}) 
\rc +\cdots~. 
\eea
Thus, the retarded Green's functions of the longitudinal modes are given by
\bea
\mathcal{G}_{tt} &=& - \frac{1}{g_4^2} \lb \frac{q^{2}}{i \o - \ls\frac{\tilde{\L}}{T_{H}}\rs q^2} \rb, \la{eq:ttgreen}\\
\mathcal{G}_{yy} &=&  - \frac{1}{g_4^2} \lb \frac{\o^{2}}{i \o - \ls\frac{\tilde{\L}}{T_{H}}\rs q^2}  \rb , \la{eq:yygreen}\\
\mathcal{G}_{ty} &=&  \mathcal{G}_{yt} =  - \frac{1}{g_4^2} \lb \frac{\o q}{i \o - \ls\frac{\tilde{\L}}{T_{H}}\rs q^2} \rb,
\la{eq:tygreen}
\eea
where we have re-expressed $\tilde\o$ and $\tilde{q}$ in terms of $\o$ and $q$.
From the above expression, the longitudinal modes have a quasi normal pole as we expected and
the charge diffusion constant $D$ of this quasi normal mode is
\be		\la{res:chdiffcoeff}
 D = \frac{\tilde{\L}}{T_{H}} = \frac{(-\L)}{16\pi T_H}  \frac{(4+\eta^2)^2}{4-\eta^2} .
\ee
For the AdS black brane ($\eta = 0$ and $\L = -3$), the charge diffusion constant is given by 
\be
D = \frac{3}{4\pi T_H} , 
\ee
which shows the pole structure of the quasi normal mode in the dual conformal gauge theory.  
At this point we would like to emphasize certain points. It is evident from \eq{eq:gaugefourier} and the 
dispersion relation $\o = - i D q^2$ that the quasi normal mode decays with a half-life time $t_{1/2} = \frac{1}{D \ q^2}$.  
In case of conformal gauge theory, once we fix the temperature the diffusion constant is determined uniquely.
Thus, in the hydrodynamic limit it is obvious that a quasi normal mode with a comparatively larger momentum
will decay rapidly. 
Alternatively, in a high temperature system the quasi normal mode can sustain for longer time. 
On the other hand, in the non-conformal dual gauge theory, the diffusion constant depends on the temperature 
as well as on the parameter $\eta$ and \eq{res:chdiffcoeff} clearly shows that the diffusion constant 
increases with $\eta$. Thus, we infer that when the system deviates from the conformality, the 
quasi normal mode decays more rapidly. 

The Green function of the transverse mode is
\be
\mathcal{G}_{xx} =   \frac{1}{g_4^2}  \lb i \o - \ls\frac{\tilde{\L}}{T_{H}}\rs q^2 \rb , \la{eq:xxgreen}\\
\ee
which has no pole as we expected. Interestingly, the Green function of the transverse mode \eq{eq:xxgreen}
turned out to be the inverse of the longitudinal one  up to a multiplication factor.
The real DC conductivity of this 
system can be easily read off from \eq{eq:xxgreen}
\be
\s = \lim_{\o \to 0} {\rm Re} \ls \frac{\ {\cal G}_{xx}}{i \o} \rs = \frac{1}{g_4^2} .
\ee
Note that the non-conformality does not influence the DC conductivity. 

\section{Gauge Fluctuations coupled to dilaton field}
From now on, we will investigate the charge diffusion constant and DC conductivity with a 
nontrivial dilaton coupling. As shown in \ct{Lee:2010qs}, the dilaton coupling can affect the 
dual hydrodynamics significantly. The nontrivial dilaton profile, as we shall see, gives rise to the DC 
conductivity that depends on temperature and $\eta$.
To see this, we consider Maxwell field fluctuations coupled to a dlaton on the fixed background 
geometry \eq{ansatz}  
\be
S_{MD} = -\frac{1}{4g^{2}_{4}}\int d^{4}{x} \sqrt{-g}e^{\alpha\phi} F^{\m\n}F_{\m\n}, \la{eq:dmaxwellaction} 
\ee
where $e^{\a \ph}$ is the bulk gauge coupling depending on the radial coordinate. Notice that we can apply the same 
methodology used in the previous section. Instead of giving the details of the computation, here we only state 
the necessary steps. The equations of motion for gauge fluctuations with the dilaton coupling are
\be
0=\pa_{\n}\lb\sqrt{-g} g^{\n\rho}g^{\m\sigma}e^{\alpha\phi}F_{\sigma\rho}\rb . \la{eq:dilgaugefluct}
\ee
We introduce a new parameter such that,
 \be 
\delta  = -\alpha k_0 = \frac{\eta^2}{4+\eta^2} .
 \ee
Although $\alpha$ can have any arbitrary value, here we choose a special value,
$\alpha = -\eta/2$, for more concrete calculation. Now since $0 \le \eta^2 < 4$, it automatically 
puts a bound on $\delta$ as $0 \le \delta < 1/2$. From \eq{eq:dilgaugefluct}, we get the following set of 
coupled equations for $A_{t}$ and $A_{y}$ 
\bea
0 &=& b^2 \tilde{\o} A'_{t} + g \tilde{q} A'_{y} , \la{eq:dArr}\\
0 &=&b^2 A''_{t} +2bb' A'_{t} + b^2 A'_{t} \ls\frac{\delta}{r}\rs - \frac{T^{2}_{H}}{g}(\tilde{\o}\tilde{q} A_{y}+ \tilde{q}^2 A_{t}) \la{eq:dAtr}\\
0 &=& g A''_{y} + g' A'_{y} +g A'_{y}\ls\frac{\delta}{r}\rs + \frac{T^{2}_{H}}{g}(\tilde{\o}\tilde{q} A_{t}+\tilde{\o}^{2}A_{y})  \la{eq:dAyr} 
\eea
 where the prime denotes derivative with respect to the radial coordinate $r$.
For the transverse mode $A_{x}$, we have the following equation
\be 
0 = A''_{x} + \frac{g'}{g}A'_{x} +A'_{x}\ls\frac{\delta}{r}\rs  + \frac{T^{2}_{H}}{g^2}\lb\tilde{\o}^2 -\tilde{q}^2\frac{g}{b^2}\rb A_{x} .
\la{eq:dAxr} 
\ee

For simplicity, we introduce new coordinate $u$ 
\be
u = 
\ls\frac{T_{H}}{\tilde{\L}_{eff}(T_{H})}\rs\frac{r^{1-(2a_{1}+\delta)}}{(2a_{1}+\delta -1)} \la{eq:ucoordinate}
\ee
with
\be
\tilde{\L}_{eff} = \tilde{\L} \ \frac{2a_{1}-1}{2a_{1}+\delta -1}\lb \frac{\tilde{\L}}{T_{H}}\ls
 \frac{4-\eta^{2}}{4+\eta^{2}}\rs\rb^{\delta\ls\frac{4+\eta^{2}}{4-\eta^{2}}\rs} ,
\ee
where $\td{\L}$ was defined in \eq{eq:lambdalambdatilde} and $\tilde{\L}_{eff}$ has a nontrivial 
dependence on temperature. The metric coefficients in $u$ coordinate become 
\bea
g(u) &=& a^{2}_{0}\lb u (2a_{1}+\delta -1)\ls\frac{\tilde{\L}_{eff}}{T_{H}}\rs\rb^{e}
\lb 1-u^{d}\rb \la{eq:gz}\\
b^{2}(u) &=& \lb u (2a_{1}+\delta -1)\ls\frac{\tilde{\L}_{eff}}{T_{H}}\rs\rb^{e} \la{eq:dbu}
\eea
with
\be
d = \frac{c}{2a_{1}+\delta -1} \ , \  e = -\frac{2a_{1}}{2a_{1}+\delta -1}. \la{eq:ddande}
\ee
Since $0 \le \eta^2 < 4$, there exists a bound on $d$ which is $2< d \le 3$ and $e$ is found to be $-2$.

Then the coupled equations for the longitudinal modes are rewritten as
\bea
0 &=&\tilde{\o} A'_{t} + F(u) \tilde{q} A'_{y} ,\la{eq:dArz}\\
0 &=&A''_{t} - \frac{\tilde{\L}_{eff}^2}{H(u)}\lb \tilde{q}\tilde{\o} A_{y} + \tilde{q}^2 A_{t} \rb , \la{eq:dAtz}\\
0 &=&A''_{y} + \frac{F'(u)}{F(u)}A'_{y} +\frac{\tilde{\L}_{eff}^2}{F(u)H(u)}\lb \tilde{\o}\tilde{q}A_{t} +
 \tilde{\o}^2 A_{y}\rb, \la{eq:dAyz}
\eea
while the differential equation of the decoupled transverse mode $A_{x}$ is given by
\be 
0=A''_{x} + \frac{F'(u)}{F(u)} A'_{x} + \frac{\tilde{\L}_{eff}^2}{F(u)H(u)}\lb \tilde{\o}^2 -
 F(u)\tilde{q}^2\rb , \la{eq:dAxz} 
\ee
where the prime implies the derivative with respect to $u$. Here, we define several new functions 
\bea
F(u) &=& \frac{g(u)}{b^{2}(u)} = a^2_{0}(1-u^d)\la{eq:dF} , \\
H(u) &=& \frac{g(u)}{B^{2}(u)} ,
\eea
where
\bea
B^{2}(u) &=& \lb u (2a_{1}+\delta-1)\ls\frac{\tilde{\L}_{eff}}{T_{H}}\rs\rb^{-\gamma} \ b^{2}(u), \\
\gamma &=& \frac{2\delta}{2a_{1}+\delta-1} = \frac{\eta^2}{2} . \la{eq:gammadelta}
\eea
Therefore $\gamma$ is bounded: $0 \le \gamma < 2$.

Again the details of the vector fluctuation including the dilaton field can be found in Appendix B.

\subsection{Charge diffusion constant and Conductivity}

Near the asymptotic boundary $u=0$, the solutions we found for the longitudinal modes 
\eq{eq:dAyprime} imply the following 
relationship between the radial derivatives of the fields and their boundary values,
\bea 
A'_{t} &=& \ls\frac{\tilde{\L}_{eff}^2}{a^{2}_{0}} \frac{ \ls\tilde{\o}\tilde{q} A^{0}_{y} + \tilde{q}^2 A^{0}_{y}\rs}{\lb (2a_{1}+\delta-1)\ls\frac{\tilde{\L}_{eff}}{T_{H}}\rs\rb^{\gamma}}\rs \frac{u^{1-\gamma}}{1-\gamma}  + \ \frac{\tilde{\o}\tilde{q} A^{0}_{y} + \tilde{q}^2 A^{0}_{t}}{\ls \frac{i\tilde{\o}}{\tilde{\L}_{eff}}\lb (2a_{1}+\delta-1)\ls\frac{\tilde{\L}_{eff}}{T_{H}}\rs\rb^{\gamma/2} -
 \tilde{q}^2\rs} \la{eq:dsolnofAtprimeboundary} , \\
A'_{y} &=& - \ls\frac{\tilde{\L}_{eff}^2}{a^{2}_{0}}\frac{ \ls\tilde{\o}\tilde{q} A^{0}_{t} + \tilde{\o}^2 A^{0}_{t}\rs}{\lb (2a_{1}+\delta-1)\ls\frac{\tilde{\L}_{eff}}{T_{H}}\rs\rb^{\gamma}}\rs \frac{u^{1-\gamma}}{1-\gamma}  - \ \frac{\tilde{\o}\tilde{q} A^{0}_{t} + \tilde{\o}^2 A^{0}_{y}}{\ls \frac{i\tilde{\o}}{\tilde{\L}_{eff}}\lb (2a_{1}+\delta-1)\ls\frac{\tilde{\L}_{eff}}{T_{H}}\rs\rb^{\gamma/2} - \tilde{q}^2\rs} .\la{eq:dAyprimeboundary}
\eea 
From equations (\ref{eq:dsolnofAtprimeboundary}) and (\ref{eq:dAyprimeboundary}) we see that 
for $0 \le \gamma < 1$ the first terms will give vanishing contribution near the boundary 
however, for  $1 <\gamma < 2$ those are divergent terms.

Similarly, near the asymptotic boundary $u=0$, the solution for the transverse mode 
$A'_{x}$ in \eq{eq:dsolnofAxz} and \eq{eq:dCx} is related to 
the boundary value of $A_{x}$ in the following way,
\bea
&& A'_{x} \nn 
\hspace{-0.5cm} &&= \frac{ A^{0}_{x} \tilde{\L}_{eff}^2}{a^{2}_{0} \lb (2a_{1}+\delta-1)\ls\frac{\tilde{\L}_{eff}}{T_{H}}\rs\rb^{\gamma}} \lb \frac{i\tilde{\o}}{\tilde{\L}_{eff}}\lc (2a_{1}+\delta-1)\ls\frac{\tilde{\L}_{eff}}{T_{H}}\rs\rc^{\gamma/2} -\frac{\tilde{q}^2}{1-\gamma}  
\ls 1 - \ u^{1-\gamma} \rs \rb .
\la{eq:dAxprimeboundary}
\eea
 Here, we see that for $0 \le \gamma < 1$ the first term in (\ref{eq:dAxprimeboundary}) will give vanishing 
 contribution near the boundary $u=0$ however, $1 <\gamma < 2$ that is a divergent piece, which may be cancelled by 
 adding appropriate counter term following the holographic renormalization scheme.

\begin{figure}
\begin{center}
\vspace{-1cm}
\hspace{-0.5cm}
\subfigure[]{ \includegraphics[angle=0,width=0.45\textwidth]{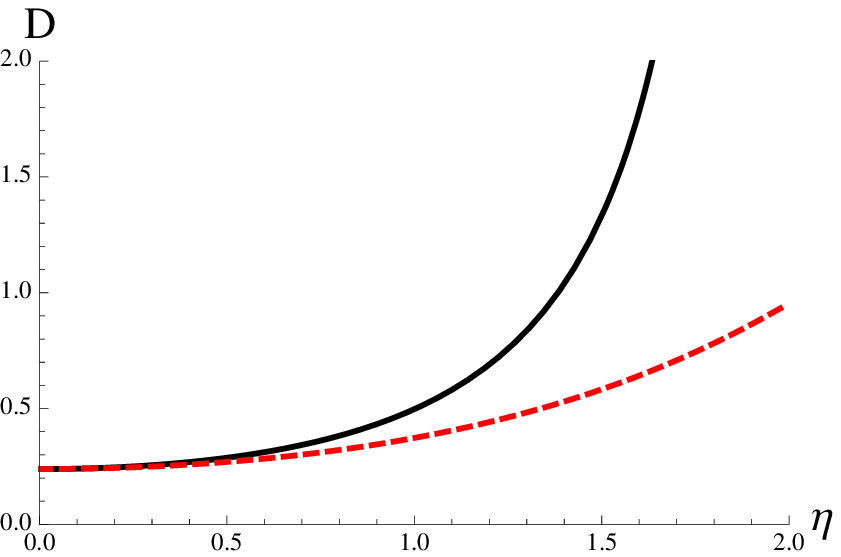}}
\hspace{-0cm}
\subfigure[]{ \includegraphics[angle=0,width=0.45\textwidth]{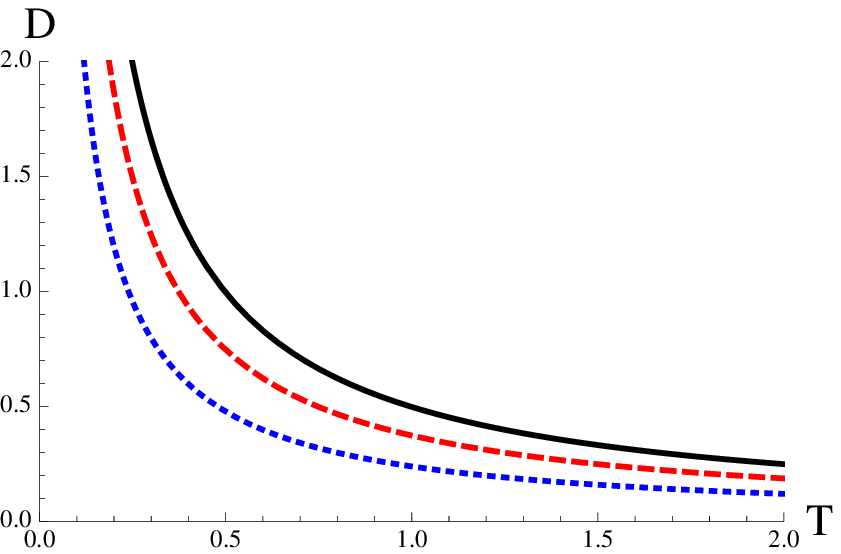}}
\vspace{-0cm} \\
\caption{\small Charge diffusion constants: (a) $\eta$ dependence and (b) temperature dependence.
The dashed and solid curve implies with or without a non-trivial dialton coupling, respectively (The $AdS_4$ result is represented as a dotted curve).}
\label{chargediff}
\end{center}
\end{figure}

Finally applying the prescription formulated in Section 3.2 one finds the non-vanishing components of the Green's function to be
\bea
\mathcal{G}_{tt} &=& - \frac{1}{g^{2}_{4}}\frac{q^2}{\ls i\o\lb (2a_{1}+\delta-1)\ls\frac{\tilde{\L}_{eff}}{T_{H}}\rs\rb^{\gamma/2} -
\ls\frac{\tilde{\L}_{eff}}{T_{H}}\rs q^2\rs} ,\la{ttgreen} \\
 \mathcal{G}_{yy} &=& - \frac{1}{g^{2}_{4}}\frac{\o^2}{\ls i\o\lb (2a_{1}+\delta-1)\ls\frac{\tilde{\L}_{eff}}{T_{H}}\rs\rb^{\gamma/2} -
\ls\frac{\tilde{\L}_{eff}}{T_{H}}\rs q^2\rs} ,\la{yygreen} \\
 \mathcal{G}_{ty}  &=&  \mathcal{G}_{yt} = - \frac{1}{g^{2}_{4}}\frac{\o q}{\ls i\o\lb (2a_{1}+\delta-1)\ls\frac{\tilde{\L}_{eff}}{T_{H}}\rs\rb^{\gamma/2} -
\ls\frac{\tilde{\L}_{eff}}{T_{H}}\rs q^2\rs} ,\la{ytgreen} \\
\mathcal{G}_{xx} &=&  \frac{1}{g^{2}_{4}}\lb
\frac{i\o}{ \lb (2a_{1}+\delta-1)\ls\frac{\tilde{\L}_{eff}}{T_{H}}\rs\rb^{\gamma/2}} -
\frac{q^2}{ \ls 1-\gamma \rs \lb (2a_{1}+\delta-1) \ls \frac{\tilde{\L}_{eff}}{T_{H}}\rs 
\rb^{\gamma}} \ls \frac{\tilde{\L}_{eff}}{T_{H}} \rs\rb  .\la{xxgreen}
\eea
In this case, due to the dilaton coupling, the Green function of the transverse mode is not exactly the 
inverse of that of the longitudinal one. From the above expression, we can easily read off the diffusion constant $D$, to be
\be		\la{res:cdc}
 D = \frac{(- \L)  }{16 \pi T_H}  \frac{ (4 + \eta^2)^2}{4}.
\ee
Since the charge diffusion constant in \eq{res:cdc} is almost similar to that in \eq{res:chdiffcoeff}, 
the quasi normal modes possess similar qualitative features. In Fig. \ref{chargediff}, we compare the charge diffusion
constants of the quasi normal modes with and without a nontrivial dilaton coupling.

\begin{figure}
\begin{center}
\vspace{-1cm}
\hspace{-0.5cm}
\subfigure[]{ \includegraphics[angle=0,width=0.45\textwidth]{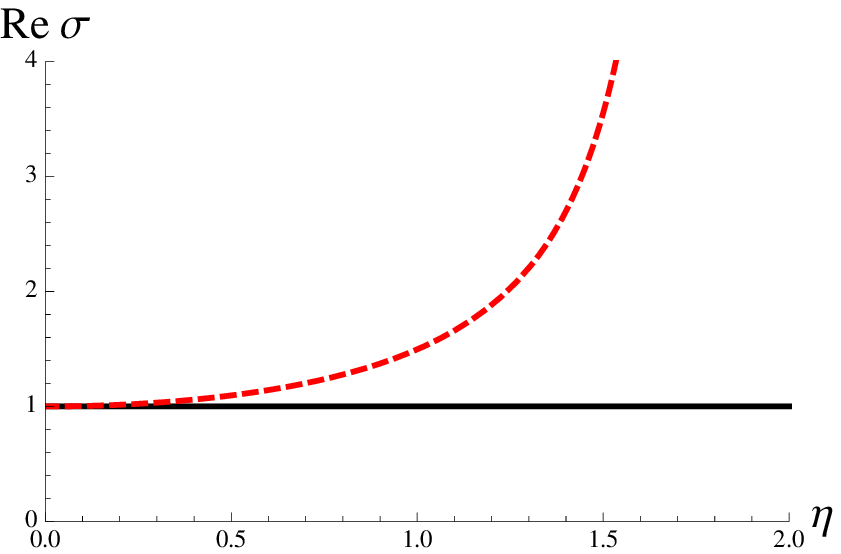}}
\hspace{-0cm}
\subfigure[]{ \includegraphics[angle=0,width=0.45\textwidth]{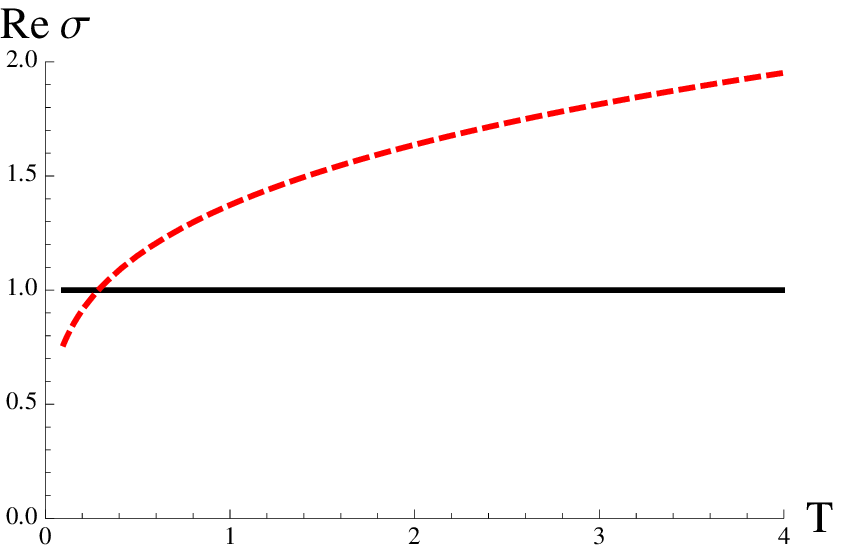}}
\vspace{-0cm} \\
\caption{\small Real conductivities: (a) $\eta$ dependence and (b) temperature dependence.
The dashed and solid curve implies with or without a dialton coupling respectively.}
\label{conduct}
\end{center}
\end{figure}

From the retarded Green function of the transverse mode, the real DC conductivity is given 
by 
\be		\la{res:condd}
\s = \frac{1}{g^{2}_{4}}\lb \frac{16 \pi}{(-\L) (4 + \eta^2)} \rb^{\frac{\eta^2}{4-\eta^2}} \ T_H^{\frac{\eta^2}{4-\eta^2}} ,
\ee
which shows a significant change in the behaviour from that of the previous one. 
The nontrivial dilaton coupling gives rise to a conductivity that obeys a power law behaviour.
We can interpret this new feature from the dual gauge theory point of view as follows.  
In the condensed matter theory, the crucial fact that conductivity depends on temperature  
relies on the very nature of the charge carrier. For metals, the conductivity 
usually decreases with temperature. This can be attributed to the fact that 
the motion of electrons in the medium is significantly disturbed at high temperature. 
Unlike metals, the conductivity of common electrolytes, where the charge carriers are ions, 
increases with temperature. Another example illustrating similar behaviour is 
a conductive polymer (see Fig.1 in \ct{Kaynak}). Therefore, it is interesting to investigate 
further the Einstein-dilaton theory and its dual and compare the results to those of the above materials, which would be instrumental in understanding the physics of such real materials. 
In Fig. \ref{conduct}, we plot the real conductivity depending on $\eta$ and temperature with and 
without a nontrivial dilaton coupling.

\section{Discussion}
We have studied the black hole solution of the Einstein-dilaton theory with
a Liouville potential. This solution having a non-zero scalar profile modifies the asymptotic
geometry from AdS to an warped space. The warped geometry with one parameter $\eta$ 
preserves the $ISO(1,2)$ isometry. This can be identified with the Poincare symmetry group of 
the dual gauge theory, once the gauge/gravity duality is applied to this background. 
The Einstein-dilaton theory with a Liouville type potential can be obtained  from the string theory
after suitable compactification. Hence, we believe that the gauge/gravity duality, 
which is nothing but the closed/open string duality in disguise, would be still applicable to 
our background. It is obvious that, when $\eta$ becomes zero, the warped geometry is reduced 
to the AdS space with a larger conformal symmetry group $SO(2,3)$ of the dual gauge theory.
Further, we have investigated the type of the gauge theory which appears to be the 
dual to the warped geometry and found that the Liouville parameter $\eta$ is related to 
the equation of state parameter, $w = (4-\eta^{2})/8 $,  of the non-conformal matter. We have 
also shown that the non-conformal gauge theory is thermodynamically stable only for the parameter 
range $0 \le \eta^{2} < 4$, where the specific heat of the dual system is positive.

 Next, in order to understand the properties of the dual theory further, 
 we have studied the linear response of the vector fluctuations in the warped geometry. 
As is evident from our calculations, due to the choice of the momentum along
$y$ direction we end up getting a set of coupled differential equations for $A_{t}$ and $A_{y}$
(longitudinal modes), while the transverse mode $A_{x}$ propagates independently.
We solved the resultant equations perturbatively in the hydrodynamic limit. 
When the gauge fields couple only to gravity, the expressions for the charged diffusion 
constant and the DC conductivity take the form similar to that of AdS dual gauge theory
apart from the $\eta$ dependence, which can be traced back to the fact that the dual gauge theory
is non-conformal. We observed that, the $\eta$ dependence increases the charge diffusion constant
compared to its AdS counterpart. In a nutshell, the quasi normal mode in the non-conformal medium 
has shorter life-time than that of the conformal case. We have also found that the longitudinal modes 
have a quasi normal pole. The DC conductivity, computed from the Green's function of the transverse mode 
is proportional to the bulk gauge coupling. Since the bulk gauge coupling is constant, there is no 
significant difference between the DC conductivities of the conformal and non-conformal matter. 

Finally, we considered the gauge fluctuations coupled with the dilaton. This kind of nontrivial dilaton 
coupling provides a peculiar physical aspect to the dual gauge theory such as the strange metallic behavior. 
Here for definiteness, we chose a specific value, $\alpha = -\eta/2$. With this choice, the charge diffusion 
constant has a similar form to that of the previous one with some modifications. For a fixed non-conformality, 
the diffusion constant is smaller than the one, obtained from the dilaton free gauge fluctuation. 
This also leads to the fact that the quasi normal mode with the dilaton coupling survives longer than that 
of the free one. We have shown that the effective bulk gauge coupling depending nontrivially on the radial coordinate
can change the behaviour of the DC conductivity, dramatically. The DC conductivity of the system increases with temperature, 
which is a typical aspect commonly found in electrolytes. Another example with such a temperature dependence can be found in conductive polymers such as
polypyrrole films \ct{Kaynak}.  Thus, it would be interesting to investigate at length, the interplay between the 
dual gauge theory of the warped geometry and the condensed matter system. We also would like to explore the paradigm 
by including the metric fluctuation and thus determine  other hydrodynamical quantities like shear viscosity etc.  
We hope to report on these issues in near future. 
\vspace{1cm}

{\bf Acknowledgement}

C. Park would like to thank E. Kiritsis, R. Meyer and S. J. Sin for the valuable discussions.
This work was supported by the National Research Foundation of Korea(NRF) grant funded by
the Korea government(MEST) through the Center for Quantum Spacetime(CQUeST) of Sogang
University with grant number 2005-0049409. C. Park was also supported by Basic Science Research Program through the National Research Foundation of Korea(NRF) funded by the Ministry of Education, Science and Technology(2010-0022369).

\vspace{1cm}

\appendix

\renewcommand{\theequation}{A.\arabic{equation}}
\setcounter{equation}{0}
\section{Vector fluctuation without the dilaton coupling}

\subsection{Longitudinal modes: $A_t (z)$ and $A_y (z)$}

The three equations of motion for the longitudinal modes, \eq{eq:Arz}, \eq{eq:Atz} and \eq{eq:Ayz}, are not independent because combining two of them yields the rest. By using \eq{eq:Arz} and \eq{eq:Atz} we get the following
third order differential equation
\be
0 = A'''_{t} + \frac{F'(z)}{F(z)}A''_{t} + \frac{\tilde{\L}^2}{F^2(z)}\lb \tilde{\o}^2 -
 F(z)\tilde{q}^2\rb A'_{t} . \la{eq:Atprime}
\ee
Since $F(z)$ vanishes at $z=1$, the solution of $A'_t$ should have a singular part at the horizon. 
If we take the following ansatz
\be
A'_{t}(z) = (1-z^d)^{\n} G(z) , \la{eq:longansatz}
\ee
then the unknown function $G(z)$ should be regular and at the same time independent of $\td{\o}$ and $\td{q}^2$ at the horizon,
the power $\n$ can be exactly determined by the singular structure at the horizon to be,
\be
\n = \pm \frac{i \tilde{\o}\tilde{\L}}{d a^{2}_{0}} \la{eq:nu} .
\ee
Since the black hole absorbs all kinds of field, there is no outgoing modes at the horizon.
Therefore, it is natural to impose the incoming boundary condition to $A'_t$ at the horizon,
which breaks the unitarity of the solution. In \eq{eq:nu}, the plus and minus signs correspond to  
the outgoing or incoming mode, respectively and hence we have to choose the minus sign.
 
Then, the equation governing the unknown function $G(z)$ becomes 
\bea
0 &=& G''(z) - \frac{1}{1-z^d} \  d (1+2\n)  z^{d-1} \ G'(z) - \frac{1}{1-z^d} \lb d(d-1) \n
z^{d-2}+\frac{\tilde{\L}^2}{a^{2}_{0}}\tilde{q}^{2} \rb \ G(z) \nn
&&+\frac{1}{\ls 1-z^d \rs^2 }  \ \lb d^{2}\n^2 z^{2(d-1)}  
+ \frac{\tilde{\L}^2}{a^4_{0}}\tilde{\o}^2\rb\ G(z) . \la{eq:diffeqGz}
\eea
Solving the above equation analytically is almost impossible, so
we have to resort either to the numerical method or take an appropriate approximation.
Here, we will consider the hydrodynamic limit wherein, $\tilde{\o} << 1$ and $\tilde{q}^2 << 1$,
and expand $G(z)$ in the powers of $\tilde{\o}$ and $\tilde{q}^2$ as:
\be
G(z) = G_{0}(z) + \tilde{\o} G_{1}(z) + \tilde{q}^2 G_{2}(z) + \tilde{\o}\tilde{q}^2 G_{3}(z) + \cdots \la{eq:seriesGz}
\ee
After substituting \eq{eq:seriesGz} into \eq{eq:diffeqGz}, we will determine $G(z)$ up to
leading orders in $\tilde{\o}$ and $\tilde{q}^2$.\\

At the zeroth order, \eq{eq:diffeqGz} reads
\be
0 = G''_{0}(z) - \frac{d z^{d-1}}{1-z^d} \ G'_{0}(z) \la{eq:diffG0} ,
\ee
whose solution is given by
\be
G_{0}(z)  = C_{0} \ z \  {_2}F_{1} \lb 1,\frac{1}{d},1+\frac{1}{d},z^d \rb + C ,
\ee
where $C_0$ and $C$ are two integration constants. Notice that the hypergeometric function diverges at the horizon $z=1$.
Since $G_{0}(z)$ at horizon should be regular as mentioned earlier, the divergence of the 
hypergeometric function should be removed by setting $C_{0}=0$. Hence, we have
\be
G_{0}(z) = C \la{eq:G0}  \la{res:zerothorder},
\ee
where $C$ is an undetermined constant and will be fixed later by imposing another boundary condition at the boundary. 

At $\tilde{\o}$ order, we arrive at the following differential eqn.
\be
\lb (1-z^d)G'_{1}(z)\rb^{'} = -i C \frac{(d-1)\tilde{\L}}{a^{2}_{0}}z^{d-2}  ,
\ee
where the result in \eq{res:zerothorder} was used.
The solution of the above equation is given by
\be
G_{1}(z) = C_{4}  + \frac{C_{3}  z^{1+d}}{(d+1)}  \ {_2}F_{1} \lb 1,1+\frac{1}{d},2+\frac{1}{d},z^d \rb
+ \lb C_{3}  z + i C \ls\frac{\tilde{\L}}{da^{2}_{0}}\rs\ln(1-z^d)\rb \la{eq:G1withconst1}  .
\ee
Notice that the higher order solutions like $G_1 (z)$, $G_2 (z)$, ..., must vanish  at the horizon in 
order to give a constant number. The hypergeometric function and the last term in 
\eq{eq:G1withconst1} contain divergent terms at the horizon. Thus, the integration constant $C_3$ 
should be related to $C$ in order to remove the divergence.
The remaining constant $C_{4}$ can be also determined in terms of $C$ due to the vanishing of $G_{1} (z)$ at 
the horizon \cite{Policastro:2002se, Policastro:2001yb}
\bea
C_{4} = i C \ls\frac{\tilde{\L}}{d a^{2}_{0}}\rs \lb d- EG - PG(0,1+1/d)\rb , \la{eq:C4}
\eea
where $EG$ is Eulergamma number and PG(a,b) denotes the $a^{th}$ derivative of digamma function $\psi(b)$.
As a result, $G_{1}(z)$ is exactly determined in terms of $C$ 
\be
 G_{1}(z)= \frac{iC \tilde{\L}}{a^{2}_{0}} \tilde{G}_{1}(z), \la{eq:G1z} 
\ee
where
\bea
\tilde{G}_{1}(z)&=&\frac{1}{d(d+1)}\lb  z^{1+d}d \ {_2}F_{1} \lb 1,1+\frac{1}{d},2+\frac{1}{d},z^d \rb \right. \nn
&& \quad \quad \quad \left. \frac{}{} + (1+d)\lc (d \ z +\ln(1-z^d)) -  \ls d- EG - PG(0,1+\frac{1}{d}) \rs \rc \rb .
\la{eq:G1tilde}
\eea

Following the same procedure, we can also fix $G_2 (z)$ in terms of $C$ at 
$\td{q}^2$ order. The solution $G_2 (z)$ is given by
\bea
G_{2}(z) =  C\ls\frac{\tilde{\L}^2}{2a^{2}_{0}}\rs \ \frac{z^{2}}{2} \ {_2}F_{1} \lb 1,\frac{2}{d},1+\frac{2}{d},z^{d} \rb
+ C_{5}  \ z \ {_2}F_{1} \lb 1,\frac{1}{d},1+\frac{1}{d},z^{d} \rb +C_{6} . 
\eea
The regularity and vanishing conditions of $G_2$ at the horizon fix two integration constants
$C_5$ and $C_6$ in terms of $C$
\bea
C_{5} &=& -C\frac{\tilde{\L}^2}{a^2_{0}} \la{eq:C5}\\
C_{6} &=& -C\frac{\tilde{\L}^2}{a^2_{0}d}\lb PG(0,1/d)-PG(0,2/d)\rb . \la{eq:C6}
\eea
Finally, we get the expression for $G_{2}(z)$ as
\be
G_{2}(z)=C\frac{\tilde{\L}^2}{a^2_{0}}\tilde{G}_{2}(z)\la{eq:G2z}
\ee
with
\bea
\tilde{G}_{2}(z) &=&\frac{z^{2}}{2} \ {_2}F_{1} \lb 1,\frac{2}{d},1+\frac{2}{d},z^{d} \rb -
 z \ {_2}F_{1} \lb 1,\frac{1}{d},1+\frac{1}{d},z^{d} \rb  \nn
&& - \frac{1}{d}  \ls PG(0,1/d)-PG(0,2/d  )\rs .
 \la{eq:G2tilde}
\eea

After evaluating $A''_{t}(z)$ from the above and substituting it into \eq{eq:Atz}, we obtain
\bea  \la{sol:longit}
A_{y}(z) + \frac{\tilde{q}}{\tilde{\o}} A_{t}(z) &=&\frac{C (1-z^d)^\n a^{2}_{0}}{\tilde{\o} \tilde{q} \tilde{\L}^2}\lb -d \n z^{d-1}\ls 1+\frac{i \tilde{\L}\tilde{\o}}{a^{2}_{0}} {\tilde G}_{1}(z) + \frac{\tilde{\L}^2\tilde{q}^{2}}{a^{2}_{0}}{\tilde G}_{2}(z)\rs \rp \nn
&& \qquad \qquad \qquad \quad \lp + (1-z^d)\ls \frac{i \tilde{\L}\tilde{\o}}{a^{2}_{0}}{\tilde G}'_{1}(z)+ \frac{\tilde{\L}^2 \tilde{q}^{2}}{a^{2}_{0}}{\tilde G}'_{2}(z)\rs\rb .\la{eq:AtplusAy}
\eea
where $\n$ is given in \eq{eq:nu}. To fix the overall 
integration constant $C$ we impose the Dirichlet boundary condition for $A_{t}$ and $A_{z}$ at the boundary \cite{Policastro:2002se}
\be
\lim_{z\rightarrow 0} A_{t}(z) = A^{0}_{t}   \quad {\rm and}  \quad  \lim_{z\rightarrow 0} A_{y}(z) = A^{0}_{y}. \la{eq:bcAtAy}
\ee
Rewriting $C$ in terms of the boundary values of $A_t (z)$ and $A_y (z)$, we obtain
\be
 C = \frac{\tilde{\o}\tilde{q} A^{0}_{y} + \tilde{q}^2 A^{0}_{t}}{\ls i\frac{\tilde{\o}}{\tilde{\L}}-
 \tilde{q}^2\rs  } . \la{eq:C}
\ee
Thus, the solutions up to the $\td{\o}$ and $\td{q}^2$ order are given by
\bea 
A'_{t}(z) &=& \frac{\tilde{\o}\tilde{q} A^{0}_{y} + \tilde{q}^2 A^{0}_{t}}{\ls i\frac{\tilde{\o}}{\tilde{\L}}-
 \tilde{q}^2\rs} \ (1-z^d)^{\n} \lb 1+ \frac{i\tilde{\o}\tilde{\L}}{a^2_{0}}\td{G}_{1} (z) + \frac{\tilde{q}^2\tilde{\L}^2}{a^2_{0}}
 \td{G}_{2} (z) \rb ,  \la{eq:solnofAtprime}\\
A'_{y}(z) &=& - \frac{\tilde{\o}}{a^{2}_{0} \tilde{q}} \ \frac{A'_{t}(z)}{(1-z^d)} ,\la{eq:solnofAyprime}
\eea 
where \eq{eq:Arz} was used in the last equation.

\subsection{Transverse mode: $A_{x}(z)$ }

We solve the equation for the transverse mode \eq{eq:Axz}, which is completely decoupled form longitudinal modes. A comparison between 
\eq{eq:Axz} and \eq{eq:Atprime} immediately shows that
the differential equation for the transverse mode is exactly
the same as that of $A'_{t}(z)$. Therefore, without any calculation, we can find the solution of \eq{eq:Axz}.
 \be
 A_{x}(z) = C_{x}(1-z^d)^{\n} \lb 1+ \frac{i\tilde{\o}\tilde{\L}}{a^{2}_{0}}\tilde{G}_{1}(z) + \frac{\tilde{q}^2\tilde{\L}^2}{a^{2}_{0}}
 \tilde{G}_{2}(z) \rb  + \cdots  ,\la{eq:solnofAxz}
 \ee
where $\tilde{G}_{1}(z)$ and $\tilde{G}_{2}(z)$ are given by (\ref{eq:G1tilde}) and (\ref{eq:G2tilde}), respectively.

The ellipsis means higher order terms. However, the integration constant $C_x$ is now different from the previous one.
To determine it, we again impose the Dirichlet boundary condition  
\be
\lim_{z\rightarrow 0}A_{x}(z) = A^{0}_{x} . \la{eq:bcAx}
\ee
Then, $C_x$ can be determined in terms of the boundary value of $A_x (z)$ as
\be
 C_{x}=A^{0}_{x}\lb 1- \frac{i\tilde{\o}\tilde{\L}}{a^2_{0} d} \lb d-EG-PG(0,1+1/d)\rb
- \frac{\tilde{q}^2\tilde{\L}^2}{d a^{2}_{0}}\lb PG(0,1/d) - PG(0,2/d))\rb\rb^{-1}  . \la{eq:Cx}
\ee
Thus, $A_{x}(z)$ is determined upto leading orders in $\tilde{\o}$ and $\tilde{q}^2$.

\renewcommand{\theequation}{B.\arabic{equation}}
\setcounter{equation}{0}
\section{Vector fluctuation coupled to the dilaton}

\subsection{Longitudinal modes: $A_{t}(u)$ and $A_{y}(u)$} 

Since the equations, \eq{eq:dArz}, \eq{eq:dAtz} and \eq{eq:dAyz}, are not independent we differentiate 
\eq{eq:dAtz} with respect to $u$ and then substitute it into \eq{eq:dArz} to obtain a third order 
differential equation for $A_{t}$
\be
0=A'''_{t} + \frac{H'(u)}{H(u)}A''_{t} + \frac{\tilde{\L}_{eff}^2}{H(u)F(u)}\lb \tilde{\o}^2 -
 F(u)\tilde{q}^2\rb A'_{t}. \la{eq:dAtprime} 
\ee
Note that since $H(u)$ vanishes at $u=1$, the above differential equation has a singular point at $u=1$. 
We therefore consider the following ansatz
\be 
A'_{t}(u)  = (1-u^d)^{\n} \chi(u). \la{eq:dAtansatz}
\ee
where the unknown function $\chi(u)$ is regular and at the same time independent of $\tilde{\o}$
and $\tilde{q}^{2}$ at the horizon $(u=1)$. Plugging the above ansatz, \eq{eq:dAtprime} becomes
\bea
0 &=&\lb 1-u^d\rb^\n \chi''(u) +\frac{\gamma}{z}\lb 1-u^d\rb^\n\chi'(u) - d\lb 1-u^d\rb^{\n-1} u^{d-1}\chi'(u)(1+2\n) \nn
&&-\lb 1-u^d\rb^{\n -1}\chi(u)\lb d(d-1+ \gamma)\n
u^{d-2}+\frac{\tilde{\L}_{eff}^2\tilde{q}^{2}}{a^{2}_{0}} \ \frac{1}{u^{\gamma}\lb (2a_{1}+\delta-1)\ls\frac{\tilde{\L}_{eff}}{T_{H}}\rs\rb^{\gamma}}\rb \nn
&&+\lb 1-u^d \rb^{\n -2}\chi(u)\lb d^{2}\n^2 u^{2(d-1)}  + \frac{\tilde{\L}_{eff}^2\tilde{\o}^{2}}{a^{4}_{0}} \ \frac{1}{u^{2\gamma}\lb (2a_{1}+\delta-1)\ls\frac{\tilde{\L}_{eff}}{T_{H}}\rs\rb^{\gamma}}\rb . \la{eq:ddiffeqGz}
\eea
As computed in the previous section, the index $\n$ can be determined from the singular structure at 
the horizon by solving the indicial equation and keeping in mind the fact that there are no outgoing modes 
at the horizon. Thus we get 
\be
\n = - \frac{i \tilde{\o}\tilde{\L}_{eff}}{d a^{2}_{0}} \ \frac{1}{\lb (2a_{1}+\delta-1)\ls\frac{\tilde{\L}_{eff}}
{T_{H}}\rs\rb^{\gamma/2}}  .\la{eq:dnu}
\ee 
We can solve \eq{eq:ddiffeqGz} perturbatively by considering the hydrodynamic limit and expanding $\chi(u)$ in powers of $\tilde{\o}$
and $\tilde{q}^{2}$, as
\be 
\chi(u) = \chi_{0}(u) + \tilde{\o} \chi_{1}(u) + \tilde{q}^2 \chi_{2}(u) + \tilde{\o}\tilde{q}^2 \chi_{3}(u) + \cdots . \la{eq:dseriesGz}
\ee 
Following similar steps described in the previous section, we can find 
\be 
\chi_{0}(u) = C  \la{eq:dG0}
\ee
where $C$ remains to be determined. The next order solutions, $\chi_1$ and $\chi_2$, can also be determined by imposing the
regularity and vanishing conditions at the horizon in the following form:
\bea
 \chi_{1}(u) &=& i C \ls\frac{\tilde{\L}_{eff}}{a^{2}_{0}}\frac{1}{\lb (2a_{1}+\delta-1)\ls\frac{\tilde{\L}_{eff}}{T_{H}}\rs\rb^{\gamma/2}}\rs \tilde{\chi}_{1}(u), \la{eq:dG1z} \nn
 \chi_{2}(u) &=& C \ls\frac{\tilde{\L}_{eff}^2}{a^{2}_{0}}\frac{1}{\lb (2a_{1}+\delta-1)\ls\frac{\tilde{\L}_{eff}}{T_{H}}\rs\rb^{\gamma}}\rs\tilde{\chi}_{2}(u) \la{eq:dG2z} ,
\eea
where
\bea
\tilde{\chi}_{1}(u) &=& \frac{u^{1-\gamma}}{1-\gamma} \  {_2}F_{1} \lb 1,\frac{1-\gamma}{d},1+\frac{1-\gamma}{d},u^d \rb 
 +\frac{1}{d}\lb \ln(1-u^d)+ EG + PG(0,\frac{1-\gamma}{d})\rb \la{eq:dG1tilde}  , \nn
 \tilde{\chi}_{2}(u)&=&\frac{u^{2-\gamma}}{2-\gamma} \ {_2}F_{1} \lb 1,\frac{2-\gamma}{d},1+\frac{2-\gamma}{d},u^{d} \rb - \frac{u^{1-\gamma}}{1-\gamma} \ {_2}F_{1} \lb 1,\frac{1-\gamma}{d},1+\frac{1-\gamma}{d},u^{d} \rb  \nn
&& - \frac{1}{d}\lb PG \ls 0,\frac{1-\gamma}{d} \rs -PG \ls 0,\frac{2-\gamma}{d} \rs \rb  .
 \la{eq:dG2tilde}
\eea
Here, notice that there exists a singular point at $\gamma = 1$ but as will be shown, the 
interesting physical quantities like the charge diffusion constant and the real DC conductivity, 
are well behaved, despite this singularity.

After imposing the Dirichlet boundary condition, the overall integration constant $C$ can be fixed 
(up to orders of $\tilde{\o}$ and $\tilde{q}^2$) in terms of the boundary values $A^{0}_{t}$ and $A^0_y$, 
\be
 C = \frac{\tilde{\o}\tilde{q} A^{0}_{y} + \tilde{q}^2 A^{0}_{t}}{\ls \frac{i\tilde{\o}}{\tilde{\L}_{eff}}\lb (2a_{1}+\delta-1)\ls\frac{\tilde{\L}_{eff}}{T_{H}}\rs\rb^{\gamma/2} -
 \tilde{q}^2\rs } .\la{eq:dC} 
\ee
After all the dusts get settled, one can write the final expressions for $A'_{t}(u)$ and $A'_{y}(u)$ as
\bea 
&& A'_{t}(u) = (1-u^d)^{\n}\frac{\tilde{\o}\tilde{q} A^{0}_{y} + \tilde{q}^2 A^{0}_{t}}{\ls \frac{i\tilde{\o}}{\tilde{\L}_{eff}}\lb (2a_{1}+\delta-1)\ls\frac{\tilde{\L}_{eff}}{T_{H}}\rs\rb^{\gamma/2} - \tilde{q}^2\rs}  \nn
 && \times \lb 1+ i\tilde{\o}\ls\frac{\tilde{\L}_{eff}}{a^{2}_{0}}\frac{1}{\lb (2a_{1}+\delta-1)\ls\frac{\tilde{\L}_{eff}}{T_{H}}\rs\rb^{\gamma/2}}\rs\tilde{\chi_{1}}  
  + \tilde{q}^2\ls\frac{\tilde{\L}_{eff}^2}{a^{2}_{0}}\frac{1}{\lb (2a_{1}+\delta-1)\ls\frac{\tilde{\L}_{eff}}{T_{H}}\rs\rb^{\gamma}}\rs
 \tilde{\chi_{2}} \rb \la{eq:dsolnofAtprime} , \nn
&& A'_{y}(u) = - \frac{\tilde{\o}}{\tilde{q}} \frac{1}{a^{2}_{0}(1-u^d)}A'_{t}(u) . \la{eq:dAyprime}
\eea 

\subsection{Solution for transverse mode $A_{x}(u)$}
Unlike what happened for the dilaton free fluctuations, the equation for the transverse mode $A_{x}(u)$, here is not 
the same as the one for the longitudinal mode $A'_t (u)$ in \eq{eq:dAtprime}. Hence, we need to solve \eq{eq:dAxz}  
by using the perturbative expansion in the hydrodynamic limit. The solution of the transverse mode  $A_{x}(u)$  is given by
 \bea
 A_{x}(u) &=& C_{x}(1-u^d)^{\n} \lb 1+ i\tilde{\o}\ls\frac{\tilde{\L}_{eff}}{a^{2}_{0}}\frac{1}{\lb (2a_{1}+\delta-1)\ls\frac{\tilde{\L}_{eff}}{T_{H}}\rs\rb^{\gamma/2}}\rs\tilde{\zeta_{1}} (u) \right. \nn
&& \qquad \qquad \qquad \left. + \tilde{q}^2\ls\frac{\tilde{\L}_{eff}^2}{a^{2}_{0}}\frac{1}{\lb (2a_{1}+\delta-1)\ls\frac{\tilde{\L}_{eff}}{T_{H}}\rs\rb^{\gamma}}\rs
 \tilde{\zeta_{2}} (u) \rb , \la{eq:dsolnofAxz} 
 \eea
where $\tilde{\zeta}_{1}(u)$ and $\tilde{\zeta}_{2}(u)$ are given by
 \bea
\tilde{\zeta}_{1}(u)&=& \frac{1}{d(d+1)}\lb  u^{1+d}d \ {_2}F_{1}[1,1+\frac{1}{d},2+\frac{1}{d},u^d] \right. \nn
&&\left. \frac{}{}+ (1+d)\ls(ud +\ln(1-u^d))-
(d- EG - PG(0,1+\frac{1}{d}))\rs\rb \la{eq:dzeta1tilde}  ,
\eea
 \bea
\tilde{\zeta}_{2}(u) &=& \lb\frac{u^{2-\gamma}}{(1-\gamma)(2-\gamma)} \ {_2}F_{1}[1,\frac{2-\gamma}{d},1+\frac{2-\gamma}{d},u^{d}] -
 \frac{u}{(1-\gamma)} \ {_2}F_{1}[1,\frac{1}{d},1+\frac{1}{d},u^{d}] \right. \nn
&& \left. - \frac{1}{d(1-\gamma)}\ls PG(0,\frac{1}{d})-PG(0,2-\frac{\gamma}{d})\rs\rb  .
 \la{eq:dzeta2tilde}
\eea
To determine the  overall constant $C_{x}$ we impose the Dirichlet boundary condition,
\be
\lim_{z\rightarrow 0}A_{x}(u) = A^{0}_{x} . \la{eq:dbcAx} 
\ee
Finally, we can rewrite the constant $C_{x}$ in terms of the boundary value  $A^{0}_{x}$, as
\be
 C_{x}= A^{0}_{x} \lb { 1- \frac{i\tilde{\o}\tilde{\L}_{eff} }{a^2_{0}d} \frac{\lb d-EG-PG(0,1+\frac{1}{d})\rb}{\lb (2a_{1}+\delta-1)\ls\frac{\tilde{\L}_{eff}}{T_{H}}\rs\rb^{\gamma/2} }
- \frac{\tilde{q}^2\tilde{\L}_{eff}^2}{da^{2}_{0} }\frac{\lb PG(0,\frac{1}{d}) - PG(0,\frac{2-\gamma}{d}))\rb}{(1-\gamma)\lb (2a_{1}+\delta-1)\ls\frac{\tilde{\L}_{eff}}{T_{H}}\rs\rb^{\gamma} } } \rb^{-1}\ . \la{eq:dCx}
\ee

\end{document}